\magnification=\magstep1
\overfullrule=0pt
\parskip=5pt
\font\titlefont=cmbx10 scaled\magstep1
\font\authorfont=cmr10 scaled\magstep1
\newcount\numref
\def\ref{\advance\numref by1${}^{(\the\numref)}$}
\def\cite#1{\par\noindent{\narrower\sl#1\par\noindent}}
\def\frac#1#2{\hbox{${#1\over#2}$}}
\let\phi=\varphi
\let\epsilon=\varepsilon
\centerline{\titlefont Field Theory Revisited}
\bigskip\noindent
\centerline{\authorfont C. Piron}
\bigskip\noindent
\centerline{D\'epartement de Physique Th\'eorique, 24 quai Ernest-Ansermet, CH-1211 Gen\`eve 4}
\bigskip\bigskip\noindent
{\sl Abstract\/}: Following P. A. M. Dirac's critique in
``Lectures on Quantum Field Theory'' of the usual formalism, I will
discuss the role of the time parameter to solve R. Haag's no-go
theorem on the non-equivalence of the conventional Schr\"odinger
picture and the Heisenberg picture. This is possible by first defining
in a correct way the concept of the vacuum state at a given time in
relativity. I will also discuss some consequences such as the spectral
condition.
\bigskip
We will take as our basic considerations P. A. M. Dirac's beautiful
lectures on Quantum Field Theory delivered at the Yeshiva University during
the academic year 1963 -- 1964 and published by Academic Press in
1966\ref. We will also suppose that the reader knows the abstract Fock
space construction as developed along the line introduced by D. Kastler\ref.
In his lectures cited above Dirac writes [p.6]:
\cite{We have the kets at one particular time and we may picture them as
corresponding to physical states at that time, so we retain the concept
of a physical state at a certain time. We do not have the concept of a
physical state throughout all time. The latter concept, which lies at the
basis of the Schr\"odinger picture, will not occur in the present
formulation.}
This expresses exactly or point of view, where states are states at the
actual time, and the time is a $c$-number taking a well defined value in
any possible state. This is in marked contrast to $q$-numbers, which have
in general no values in any state. In spite of appearances, the
noncommutativity of the $q$-numbers is not the very essence of such
physical objects. In fact this depends on the mathematical formalism that
you choose and its corresponding interpretation. This can be seen very well
in the Wigner representation of $p$ and $q\,$. The two objects commute with
each other but the usual rules of quantum mechanics are here completely
modified and nevertheless in any given physical state $p$ and $q$ have no values.

The concept of
a physical state throughout all time, a relativistic concept, is not a
state at all, it is a trajectory of states labeled by the time, and a solution
of a Schr\"odinger equation\ref. But as Dirac demonstrates, such solutions
do not exist for the physically justified Hamiltonian for which the
interaction is too violent at high frequencies.

To exhibit the difficulties and explain Dirac's preference for the
Heisenberg picture, let us take the very simple example of a fermionic
field (non-self-interacting and non-relativistic). Consider a one-particle
problem is non-relativistic quantum mechanics. As we know, such a system
is described by a family of Hilbert spaces labeled by some $c$-number
$\alpha\,$. In the most simple case $\alpha$ is just the time, which is
as Dirac insists a $c$-number\ref. In the general formalism\ref, each
physical observable is described by a family of self adjoint operators or
better by a the corresponding spectral families. According to G.$\,$W.$\,$
Mackey, the observables $\underline{p}$, $\underline{q}$,
$\underline{t}$ are solutions of the imprimitivity systems based on the
kinematic group of rotations and translations of $\vec{p}$, $\vec{q}$
and $t$ and we have the following representation$\,$:
\medskip\noindent
I\vskip-1.6\baselineskip{\parindent=40pt
\item{}
$\underline{q}\ {\rm:}\qquad
\{q_t=x\}$
\item{}
$\underline{p}\ {\rm:}\qquad
\{\,p_t=-i\hbar\partial_x\,\}$
\item{}
$\underline{t}\ {\rm:}\qquad
\{t_t=tI\}$
\medskip\noindent}
This representation is called the Schr\"odinger representation since the
time $t$ does not appear explicitly in the operators describing $\underline{p}$
and $\underline{q}$ and more precisely it is the representation in the $\vec{q}$
variable (with diagonal $\underline{q}$).

As we have said, the state is the state at a given time $t$ and it is
described by a ray $\phi_t(x)$ in the Hilbert space $H_t\,$, the
isomorphism which corresponds to the translation of time in the imprimitivity
relations is a passive translation which allows the comparison between
$\phi_t(x)$ in $H_t$ and $\phi_{t+\tau}(x)$ in $H_{t+\tau}\,$, it is not
the evolution which is an active translation from $t$ to $t+\tau\,$. But
such passive translations give meaning to the Schr\"odinger equation
$$i{\rm d}_t\phi_t(x)=H_t\phi_t(x)\eqno(1)$$
where $H_t$ is the Schr\"odinger operator which is self adjoint when the
evolution is induced by a unitary transformation. In this particular
case we can change from the Schr\"odinger representation to the
corresponding Heisenberg representation. For example:
\medskip\noindent
The Heisenberg representation for the free particle
\medskip\noindent
II\vskip-1.6\baselineskip{\parindent=40pt
\item{}
$\underline{q}\ {\rm:}\qquad
\{\,q_t=x+{1\over m}(-i\hbar\partial_x)t\,\}$
\item{}
$\underline{p}\ {\rm:}\qquad
\{\,p_t=-i\hbar\partial_x\}$
\item{}
$\underline{t}\ {\rm:}\qquad
\{t_t=tI\}$
\medskip\noindent}
The Heisenberg representation for the harmonic oscillator
\medskip\noindent
III\vskip-1.6\baselineskip{\parindent=40pt
\item{}
$\underline{q}\ {\rm:}\qquad
\{q_t=\cos\omega t\,x+{1\over m\omega}\sin\omega t
(-i\hbar\partial_x)\,\}$
\item{}
$\underline{p}\ {\rm:}\qquad
\{\,p_t=\cos\omega t(-i\hbar\partial_x)-m\omega\sin\omega t\,x\,\}$
\item{}
$\underline{t}\ {\rm:}\qquad
\{t_t=t\,I\}$
\medskip\noindent}
We go from the Schr\"odinger to the Heisenberg representation by a unitary
transformation labeled by $t$ but acting in each Hilbert space separately$\,$:
\medskip\noindent{\advance\baselineskip by3pt\par\centerline{\hbox{\vbox{\halign{
\hfil#\hfil
\qquad&\qquad
\hfil#\hfil\cr
$e^{{i\over\hbar}({1\over2m}p^2)t}\ $
&$e^{{i\over\hbar}({1\over2m}p^2+{m\omega\over2}q^2)t}\ $\cr
$({\rm I})\hbox to3.5truecm{\rightarrowfill}({\rm II})$
&$({\rm I})\hbox to3.5truecm{\rightarrowfill}({\rm III})$\cr
}}}}\medskip\noindent}

To be able to apply the resources of functional analysis we have to
restrict $\phi_t(x)$ for each $t$ to be in ${\cal S}(R^3)\,$, the subspace
of smooth functions of rapid decrease. But this is not enough, we have to
consider also a bigger space
${\cal H}=\int_\oplus H_td{t}$
and restrict ourselves to $\phi(t,x)\in{\cal S}(R^4)\,$. In this context,
the solutions of the Schr\"odinger equation (1) are in fact in
${\cal S}'(R^4)$ and the operator $K=i\partial_t-H_t$ acting on such
${\cal H}$ has continuous spectrum from $-\infty$ to $\infty\,$. The
Schr\"odinger solution must be interpreted as a generalised eigenvector
for the eigenvalue $0$:
$$K\phi(t,x)=0\eqno(2)$$
Consequently, in ${\cal H}$ the operator $K$ is unitarily equivalent to the
`trivial one' $i\partial_t\,$.
It is only in this bigger space ${\cal H}$ that we can give a meaning to
relativistic covariance, but we have to interpret
everything at a given time $t_0$ and as Dirac explains [p.6]:
\cite{For example, take the equation $\alpha(t_0)|A\!>\,=a|A\!>$ where $a$ is a number.
If we had that equation, we could say that $|A\!>$ represents the state
at time $t_0$ for which the dynamical variable $\alpha$ at time $t_0$
certainly has the value $a$.}
Such an interpretation of eigenvalues and eigenvectors is exactly the one
that we have always given.

Knowing the description of the one-particle states, we can define the
$N$-particle states by the Fock construction, for each value of the
$c$-number $t$ we can build the Fock space ${\cal F}(H_t)$, the space
$\oplus_n(H_t)^{\otimes n}$ after symmetrisation or antisymmetrisation,
and the corresponding creation and annihilation operators
$a^\dagger(\phi_t)$ and $a(\phi_t)\,$. By taking the direct integral
you can then construct a bigger Fock space, once again the good space in
which to implement the relativistic covariance. This gives the beginnings
of a new field theory.

Let us conclude with some remarks on such a revisited field theory.
\item{$\bullet$}
In complete analogy with the notion introduced by Dirac [p.147], at each
time $t$ we can define the vacuum $|0_t\!>$ by the condition that
$a(\phi_t)|0_t\!>\,=0$ for any $a(\phi_t)\,$. Obviously such a family $|0_t\!>$
is not unique (any $e^{\i\alpha(t)}\,|0_t\!>$ is another solution),
it is even not
normalisable being in fact in ${\cal S}'(R)\,$. Such a vacuum differs very
drastically from the usual concept and here in may cases (in particular
the examples given above) the Heisenberg and Schr\"odinger representations
are unitarily equivalent.
\item{$\bullet$}
The usual spectral condition must be modified. Here the operator of the
generator corresponding to the time-translation evolution is unbounded
in both directions but degenerate starting from some lower bound in energy.
\item{$\bullet$}
The $q$-number parts of the field are not just $q$-number Schwartz
distributions but $q$-number de Rham currents\ref, which means, among other
things, that that the test functions must be replaced by the one-particle
state functions in ${\cal S}(R^4)\,$.
\item{$\bullet$}
The usual relativistic dynamical covariance of the Poincar\'e group defines
isomorphisms of the Hilbertian structure of the global property lattice
which, in general, are implemented by non-unitary and non-irreducible
representations due to the fact that the Poincar\'e group acts also on the
$c$-number part of the field\ref.
\vfill\eject\par\noindent\parindent=5pt
\newdimen\push\setbox0=\hbox{(9)\ }\push=\wd0
{\bf REFERENCES}
\bigskip
{\parindent=\push
\item{(1)}{P. A. M. Dirac ``Lectures of Quantum Field Theory'' Academic Press, New York, 1966}
\item{(2)}{D. Kastler ``Introduction \`a l'\'electrodynamique quantique'' Dunod, Paris, 1961}
\item{}{and also ``Superquantification et alg\`ebre multilin\'eaire'' in {\sl Application de la th\'eorie des champs \`a la physique du solide} Association Vaudoise des Chercheurs en Physique, Lausanne, 1964}
\item{(3)}{C. Piron ``M\'ecanique quantique bases et applications'' Presses polytechniques et universitaires romandes, Lausanne, 1998, ch.6}
\item{(4)}{See also W. Pauli ``General Principles of Quantum Mechanics'' Springer-Verlag, New York, 1980, p.63}
\item{(5)}{C. Piron ``M\'ecanique quantique bases et applications'' Presses polytechniques et universitaires romandes, Lausanne, 1998, ch.3}
\item{(6)}{G. de Rham ``Vari\'et\'es diff\'erentiables'' Hermann, Paris, 1960, \S\S8 and 31--$\,$32}
\item{(7)}{G. C. D'Emma ``On quantization of the electromagnetic field'' {\sl Helvetica Physica Acta} {\bf53} (1980) 535-551}
\par}

\bye